\documentclass[twocolumn]{aastex701}

\newcommand{\angstrom}{\textup{\AA}}

\usepackage{amsmath}
\usepackage{bm}

\usepackage{fontawesome5}

\begin{document}

\title{Constraints on White Dwarf Hydrogen Layer Masses Using Gravitational Redshifts}

\shorttitle{Constraints on White Dwarf Hydrogen Layers}
\shortauthors{Arseneau et al.}

\correspondingauthor{Stefan M. Arseneau}
\author[0000-0002-6270-8624]{Stefan M. Arseneau}
\email{arseneau@bu.edu}
\affiliation{Department of Astronomy \& Institute for Astrophysical Research, Boston University, 725 Commonwealth Ave., Boston, MA 02215, USA}

\author[0000-0001-5941-2286]{J.~J.~Hermes}
\email{jjhermes@bu.edu}
\affiliation{Department of Astronomy \& Institute for Astrophysical Research, Boston University, 725 Commonwealth Ave., Boston, MA 02215, USA}

\author[0000-0002-3524-190X]{Maria E. Camisassa}
\email{}
\affiliation{Departament de F\'{i}sica, Universitat Polit\`{e}cnica de Catalunya, c/Esteve Terrades 5, 08860 Castelldefels, Spain}

\author[0000-0002-9090-9191]{Roberto Raddi}
\email{}
\affiliation{Departament de F\'{i}sica, Universitat Polit\`{e}cnica de Catalunya, c/Esteve Terrades 5, 08860 Castelldefels, Spain}

\author[0000-0002-4791-6724]{Evan B. Bauer}
\email{}
\affiliation{Lawrence Livermore National Laboratory, Livermore, CA 94550, USA}

\begin{abstract}

The hydrogen envelope is the outermost layer of a DA white dwarf; it makes up the entirety of the stellar photosphere, and yet its typical extent is difficult to model theoretically and remains poorly observationally constrained. As a result, hydrogen envelope mass is a substantial source of systematic uncertainty in physical properties of white dwarf, including overall masses and cooling ages. In this work, we fit a Gaussian mixture model to gravitational redshifts from high-resolution spectroscopy, paired with radius measurements from Gaia BP/RP spectra, to measure the mass-radius relation for a sample of 468 white dwarfs. Our results are in excellent agreement with the predicted mass-radius relations of state-of-the-art evolutionary models, including those from the MESA Isochrones and Stellar Tracks (MIST) library. We find that mass-radius relations such as MIST which assume a thick and mass-dependent hydrogen envelope are preferred by the observed probability density function over models which assume a constant hydrogen envelope mass. Proper treatment of the evolution of white dwarf progenitors is thus important for accurately modeling the mass-radius relation. Our results indicate that gravitational redshift measurements of large samples of white dwarfs in wide binaries are promising probes of the hydrogen envelope masses of DA white dwarfs.

\end{abstract}

\keywords{\uat{Stellar Properties}{1624} --- \uat{Stellar Structure}{1631} --- \uat{DA White Dwarfs}{348} --- \uat{Gravitation}{661} --- \uat{Stellar astronomy}{1583}}


\section{Introduction}

White dwarfs are the endpoint in stellar evolution for stars with masses less than $\approx 8-10$~$M_\odot$ \citep{1997ApJ...489..772I, 2024MNRAS.527.3602C}. These stars, which never achieve central temperatures and pressures great enough to fuse elements heavier than  oxygen or neon, are supported almost exclusively by electron degeneracy pressure \citep{1931MNRAS..91..456C, 1983bhwd.book.....S}. The electron-degenerate core, composing upwards of $99\%$ of the star by mass \citep{2017ApJ...839...11C}, is typically expected to be dominated by carbon and oxygen; however it can also be dominated by oxygen and neon or by helium for very high or low mass progenitors \citep{1993PASP..105.1373I, 2019A&A...625A..87C}. Surrounding the core are envelopes of helium and hydrogen which may be only partially electron-degenerate; the photosphere of the star can only penetrate these thin, outermost layers. Despite making up $<0.01\%$ of a white dwarf by mass, the non-degenerate hydrogen envelope makes up as much as $8\%$ of the radius of the star \citep{2022PhR...988....1S}. 

Having exhausted their capacity to undergo nuclear fusion, white dwarfs spend the rest of their lives cooling monotonically. Therefore, knowing a white dwarf's mass and effective temperature is sufficient to approximate the age of the star. This fact, paired with their ubiquity, makes white dwarfs a powerful tool for cosmochronology \citep{1987ApJ...315L..77W, 2001PASP..113..409F}. White dwarf cooling models are, however, dependent on assumptions about the structure of the star. Variations in the mass of the outermost hydrogen envelope can introduce systematic errors of $\approx 10\%$ into determinations of white dwarf masses, propagating through to cause uncertainty in the total age measurement \citep{2022ApJ...934..148H}. Moreover, the thickness of the hydrogen layer can inherently drive substantial variation in the luminosity-cooling age relation \citep{1990JRASC..84..150W, 2021NatAs...5.1170C, 2025ApJ...983..158P}. The time required to evolve to $\log(\text{L}/\text{L}_\odot) = -5$ is less than $8$~Gyr for hydrogen-deficient white dwarfs, but over $14$~Gyr for those with hydrogen-rich envelopes \citep{2017ApJ...839...11C}. 

The thickness of the hydrogen layer is thought to be set primarily by the details of stellar evolution on the asymptotic giant branch (AGB) and the post-AGB. In particular, stellar evolution models suggest that late thermal pulses on or just after the AGB can burn as much as the entire hydrogen layer \citep{2005A&A...440L...1A, 2010ApJ...717..897A}. Moreover, \cite{2015A&A...576A...9A} showed that both the treatment of convective overshoot and the occurence of a third dredge-up event during the thermally pulsing AGB stage (TP-AGB) can impact the mass of the hydrogen layer. The details of these processes are themselves uncertain: the metallicity of the progenitor star and the rate of mass loss due to stellar winds in particular are poorly understood \citep{1995A&A...297..727B, 2007A&A...464..667G, 2007A&A...476..893S, 2008ApJ...675..614P}. Other processes, such as stellar mergers \citep{2023MNRAS.520.6299K} and accretion of water-rich exocomets and asteroids \citep{2015MNRAS.450.2083R, 2017MNRAS.468..971G} can also affect the mass of the hydrogen layer as well. Because the physics of the AGB are complicated and resource-intensive to model, hydrogen-deficient white dwarf models have typically been constructed by artificially stripping the hydrogen layer to a pre-determined value \citep{2017ApJ...839...11C, 2020ApJ...901...93B}. Other models, such as those of \cite{2012MNRAS.420.1462R, 2013ApJ...779...58R}, have attempted to create realistic estimates of hydrogen layer mass, but in general the uncertain physics of the TP-AGB mean that theoretical models cannot reliably predict the hydrogen content of white dwarfs.

Typically, it is only via asteroseismic analysis of pulsating DA white dwarfs (classified as DAV) that the hydrogen envelope mass can be observationally constrained. Early asteroseismic analyses yielded conflicting results, with some finding uniform hydrogen envelope masses of $M_\text{H}/M_\text{WD}\sim 10^{-4}$ \citep{1994PhDT........29C}, and others indicating a range of allowable masses from $M_\text{H}/M_\text{WD}\sim 10^{-10} -10^{-4}$ \citep{1997ASSL..214..173F}. Ensemble asteroseismology was first performed on 83 DAVs by \cite{2009MNRAS.396.1709C} who found a mean hydrogen envelope mass of $M_\text{H}/M_\text{WD}\sim 10^{-6.3}$, but with a wide spread over five orders of magnitude. In a subsequent analysis of 44 DAVs, \cite{2012MNRAS.420.1462R} also reported evidence for a diversity of hydrogen layer masses among DAV white dwarfs with a mean value of $M_\text{H} / M_\text{WD} =(1.25\pm0.7) \times 10^{-6}$. \cite{2017ASPC..509..255C} analyzed the modes of DAVs observed by the K2 mission, finding that their clumped period distribution was most readily explained by a thick (on average $M_\text{H}/M_\text{WD} \sim 10^{-4}$) hydrogen layer; they additionally found signs that the He layer may also be thinner than previously modeled. Seismic analysis is made challenging by the fact that most DAVs show fewer modes than the number of free parameters in the structural model (many of the DAVs of \citealt{2009MNRAS.396.1709C} show only 3-5 modes), meaning that the problem of structural determination is in general under-constrained. 

\cite{2017MNRAS.470.4473P} constrained the mass of the hydrogen envelopes of 16 white dwarfs in eclipsing binary systems with M dwarfs. By combining radii measured from eclipse durations with dynamical masses determined from the well-characterized binary orbit, they made high-precision empirical measurements of the mass-radius relation. They found that most stars in their sample were consistent with models having hydrogen envelopes of $10^{-5} < M_\text{H}/M_\text{WD} < 10^{-4}$. All the objects in their sample have experienced common-envelope evolution, meaning that their hydrogen layers may not be representative of the broader white dwarf population. 

\cite{2020MNRAS.492.3540C} used photometric observations of the Balmer break as a function of cooling age to constrain the fraction of white dwarfs with very thin hydrogen layers. They found that 60 per cent of all white dwarfs have hydrogen layers in the range of $10^{-10} < M_\text{H}/M_\text{WD} < 10^{-4}$, consistent with the observed fraction of white dwarfs with hydrogen-dominated atmospheres, however their methodology is unable to probe the actual mass of the hydrogen envelope within that range. 

The compact nature of white dwarfs gives rise to another observational probe of their stellar structure: gravitational redshifts. Gravitational redshift, given by the expression
\begin{equation}
    v_\text{g} = \frac{GM}{Rc},
\end{equation}
is substantial in white dwarfs, which have typical masses of $\approx 0.6$~$M_\odot$ and typical radii of $1.4$~$R_\oplus$ \citep{2007MNRAS.375.1315K}. \cite{2010ApJ...712..585F} found a mean gravitational redshift of $32.6\pm 1.2$~km s$^{-1}$ for a sample with a mean spectroscopically derived mass of $0.575~M_\odot$. When combined with an independent measurement of stellar radius, typically made via photometric fits to the spectral energy distribution constrained by Gaia parallaxes, gravitational redshifts are a powerful tool for measuring white dwarf stellar structure: we find that variations in white dwarf hydrogen layers produce signals of $\sim 1$~km s$^{-1}$ in gravitational redshift \citep{2019MNRAS.484.2711R}. However, gravitational redshifts are challenging to measure for individual white dwarfs because their measured radial velocity consists of both their gravitational redshift and an unknown velocity component due to their space motion.

\cite{2020ApJ...899..146C} measured the mass-radius relation of white dwarfs in the Sloan Digital Sky Survey (SDSS) via gravitational redshifts by binning on their photometrically derived radii and interpreting the average radial velocity within each bin. More recently, \cite{2024ApJ...977..237C} extended this technique with a larger sample from the fifth generation of SDSS to confirm that gravitational redshift depends on the star's effective temperature. 

Gravitational redshifts of individual white dwarfs with wide binary (i.e. resolved) main sequence companions have also been measured by subtracting out the radial velocity of the main sequence companion from that of the white dwarf. \cite{2024ApJ...963...17A} did this with a sample of 135 low-resolution white dwarf spectra from SDSS, and \cite{2025A&A...695A.131R} performed a similar analysis with high-resolution white dwarf spectra. In addition, \cite{2019A&A...627L...8P, 2023MNRAS.522.3710P} measured gravitational redshifts of eight white dwarfs in the Hyades cluster by subtracting out the mean motion of that structure.

\cite{2025moi} found systematic biases of $\approx 7$~km s$^{-1}$ in radial velocities measured from comparatively low resolution SDSS spectroscopy ($R=1800$). Therefore, attempts to measure white dwarf stellar structure from gravitational redshifts must be based on spectroscopic data with sufficiently high resolution to distinguish the thermally broadened cores of white dwarf absorption lines \citep{2020A&A...638A.131N}.

In this work, we make both statistical and direct measurements of the white dwarf mass-radius relation using gravitational redshifts using a large sample of high-resolution white dwarf spectra. We compare this against realistic models of hydrogen layer thickness from grids of evolutionary sequences. In Section \ref{sec:intro} we describe our dataset of high-resolution white dwarf spectra. The methods we use to determine atmospheric parameters, radial velocities, and gravitational redshifts for our sample are presented in Section \ref{sec:fitting}. In Section \ref{sec:deconv} we describe the method by which we deconvolve our measurements from their uncertainties and infer a probability distribution for the mass-radius relation. Finally, we present our results in Section \ref{sec:results} and discuss their implications in Section \ref{sec:conclusions}.

\section{Data and Observations} \label{sec:intro}

\subsection{Single White Dwarfs From SPY}

The Type Ia Supernova Progenitor Survey (SPY; \citealt{2020A&A...638A.131N}) was a high-resolution spectroscopic survey of white dwarfs using the Ultraviolet and Visual Echelle Spectrograph (UVES) at the European Southern Observatory's $8.2$~meter Very Large Telescope (ESO VLT; \citealt{2000SPIE.4008..534D}) which collected data from 2000 to 2010. White dwarfs targeted by SPY were observed in multiple $600$~s exposures with a slit width of $2.1$~arcseconds. This configuration gives coverage of the H$\alpha$, H$\beta$, H$\gamma$ and H$\delta$ lines at a spectral resolution of $R=18,500$ ($0.35$~$\angstrom$ at H$\alpha$) which is sufficient to resolve the NLTE cores of Balmer absorption lines.

\cite{2020A&A...638A.131N} provide 1391 spectra of 643 white dwarfs. We first remove known binaries, magnetic white dwarfs, and non-DA white dwarfs using Table C2 of that paper. Additionally, we remove objects with Gaia \texttt{ruwe > 1.2} and \texttt{phot\_bp\_rp\_excess > 1.2} \citep{2025AJ....169...29S} in order to remove possible marginally resolved binaries. This leaves us with 1171 spectra of 451 white dwarfs; our radial velocities are the same as those of \cite{2025moi}. Spectra reported by \cite{2020A&A...638A.131N} are provided in air wavelengths and in the reference frame of the observer. We first use \texttt{astropy} \citep{astropy:2013, astropy:2018, astropy:2022} to correct the spectra into the heliocentric reference frame and then bring them into vacuum wavelengths using the standard corrections of \cite{1991ApJS...77..119M}.

\subsection{Common Proper Motion Systems From UVES}

We use a catalog of 49 white dwarfs with resolved main sequence binary companions from \cite{2025A&A...695A.131R}. These objects were observed with UVES following a similar observing strategy to that of the SPY survey, using multiple 600~s exposures with a slit width of $2.1$~arcseconds. They also attain a spectral resolution of $R=18,500$, sufficient to make an unbiased measurement of the star's apparent radial velocity from the Balmer lines. We remove one star from their sample: \texttt{Gaia
DR3 4005438916307756928}, which is known to have a spurious gravitational redshift measurement due to emission lines in the companion \citep{2024ApJ...963...17A, 2025A&A...695A.131R}.

\begin{figure*}
    \centering
    \includegraphics[width=\linewidth]{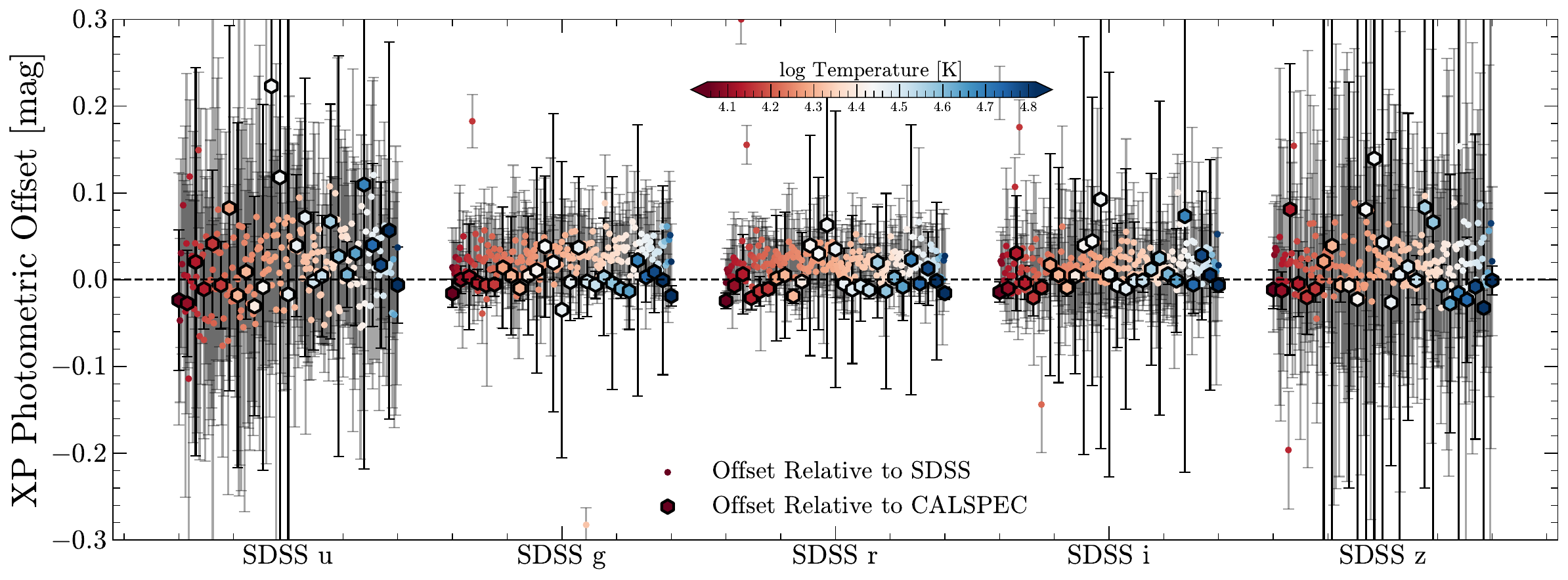}
    \caption{Difference between magnitude calculated from Gaia XP spectra and SDSS photometric bands for 203 stars from the isolated white dwarf sample for which both sources of data are available. Also shown is the difference between XP spectra and flux standard photometry for 27 white dwarfs in the CALSPEC database. Points are spaced uniformly along the horizontal axis in order of increasing effective temperature. Spacing along the horizontal axis is arbitrary and not linear across effective temperatures. We find that our photometric corrections produce photometry which is accurate to at least $3\%$ absolute flux.}
    \label{fig:calibration}
\end{figure*}

\section{White Dwarf Parameters} \label{sec:fitting}

\subsection{Photometric Radii}

We measure the radii of all stars in our sample by fitting model photometry to their observed spectral energy distributions (SEDs). We convolve 1D NLTE model spectra for a given effective temperature $T_\text{eff}$ and surface gravity $\log g$ \citep{2006ApJ...651L.137K, 2011ApJ...730..128T} into 24 synthetic photometric bands characterized by a boxcar function of width $300~\angstrom$, ranging from $3750$~$\angstrom$ to $9450$~$\angstrom$. We then scale each band by a stellar radius and the star's distance to produce a model SED:
\begin{equation}
    f_i = 4\pi H(T_\text{eff}, \log g) \left(\frac{R}{d}\right)^2 10^{-0.4 \varepsilon_{i,V} A_V}
\end{equation}
where $A_V$ is the $V$-band extinction for the source reported by \cite{2021MNRAS.508.3877G}, and $\varepsilon_{i,V}$ is a conversion factor derived by interpolating the reddening calculations of \cite{2011ApJ...737..103S} against the effective wavelength of each filter band.

The expression for the synthetic SED requires the specification of both $\log g$ and radius, however the SED does not contain enough information to constrain both of these quantities. \cite{2024ApJ...963...17A} found that the effect of $\log g$, which determines the shape of the spectral absorption lines and the shape of the Balmer jump, on convolved photometry was small. Therefore we choose the use the mass-radius relation of \cite{2020ApJ...901...93B} to compute $\log g$ for our synthetic SED calculation from $T_\text{eff}$ and radius. \cite{2025A&A...695A.131R} avoided this problem in the wide binary sample by placing a Gaussian prior on the likelihood function of the observed gravitational redshift; however this solution is not possible for isolated white dwarfs because radial velocity measurements for these objects include real space motions. We do not place a prior on gravitational redshift in order to ensure consistent systematics between the two samples.

\begin{deluxetable}{cccc}
\tablecaption{Agreement between corrected XP photometry convolved into SDSS photometric bands and actual SDSS photometry.\label{tab:sdss}}
\tablehead{
\colhead{Band} & \colhead{Median Offset [mag]} & \colhead{MAE [mag]} & \colhead{\# Stars}
}
\startdata
SDSS $u$ & 0.02 $\pm$ 0.04 & 0.04 & 201 \\
SDSS $g$ & 0.03 $\pm$ 0.03 & 0.03 & 199 \\
SDSS $r$ & 0.02 $\pm$ 0.03 & 0.02 & 203 \\
SDSS $i$ & 0.01 $\pm$ 0.05 & 0.02 & 200 \\
SDSS $z$ & 0.02 $\pm$ 0.03 & 0.03 & 203
\enddata
\tablecomments{MAE refers to mean absolute error.}
\end{deluxetable}

For each star, we use synthetic SDSS photometry convolved from low-resolution Gaia XP spectra \citep{2021A&A...652A..86C, 2023A&A...674A...2D}. XP spectra are known to suffer from systematic flux calibration issues, especially in the bluest end of the spectrograph (in particular near the SDSS $u$-band, where most white dwarfs peak in flux). \cite{2024ApJS..271...13H} derived corrections for XP spectra using flux standards from CALSPEC\footnote{\url{https://www.stsci.edu/hst/instrumentation/reference-data-for-calibration-and-tools/astronomical-catalogs/calspec}} \citep{2014PASP..126..711B, 2020AJ....160...21B, 2025AJ....169...40B} and the Hubble Next Generation Spectral Library \citep{2007ASPC..374..409H}, as well as the seventh data release of LAMOST. For each star in our sample, we first apply their corrections and then generate synthetic photometry by convolving the corrected XP spectra by the SDSS filter profiles. 

\begin{figure*}
    \centering
    \includegraphics[width=\linewidth]{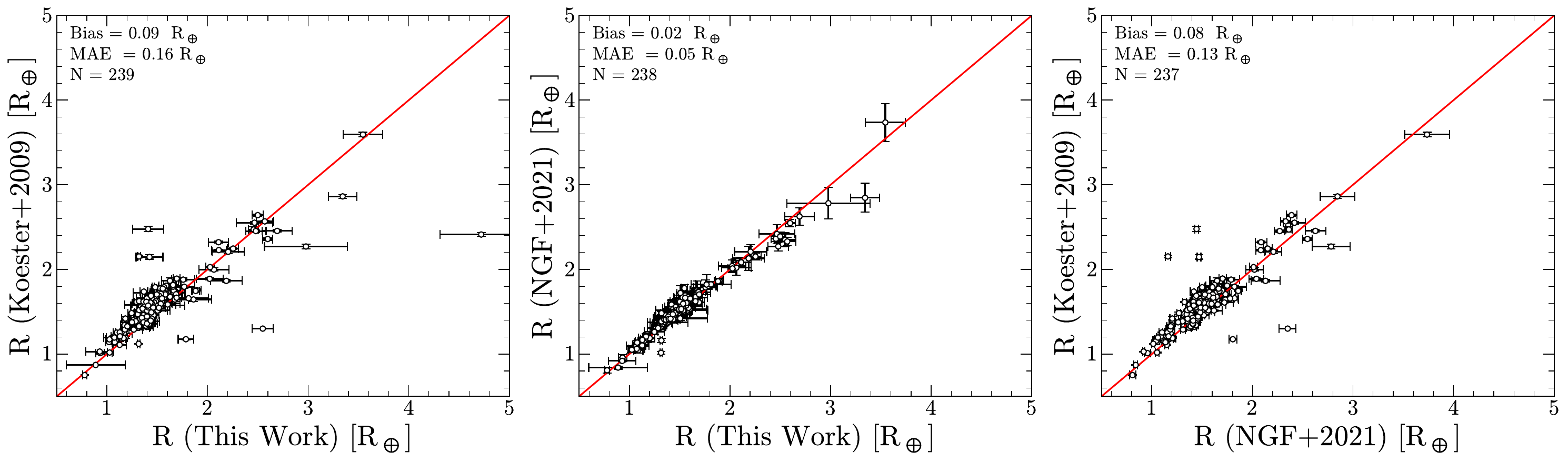}
    \caption{Measured radii (mean of the posterior distribution) from our work against the spectroscopic radius measurements of \cite{2009A&A...505..441K} and photometric radius measurements of \cite{2021MNRAS.508.3877G}. Our results are accurate up to a $1.3\%$ bias relative to the photometric measurements. For both comparison samples we transform the reported effective temperature and surface gravity into radius via the mass-radius relation of \cite{2020ApJ...901...93B}.}
    \label{fig:benchmark}
\end{figure*}

We validate the flux corrections by comparing the synthetic photometry obtained from XP spectra to photometry obtained by convolving 27 white dwarfs in the CALSPEC flux standard library with SDSS filter transmission curves. We present the results in Figure \ref{fig:calibration}. Additionally, we compare our XP photometry to real SDSS survey photometry using a sample of 203 white dwarfs with both XP spectra and SDSS photometry. For each band we neglect objects with magnitude differences between SDSS and XP photometry greater than $0.3$~mag, as these sources are likely poorly behaved (due to, for example, source contamination or photometric variability). Table \ref{tab:sdss} provides the residuals between corrected XP photometry and SDSS photometry for each filter, as well as the mean absolute error and the number of stars in the sample. Our findings indicate that the corrections work as intended, to at least 3\% absolute flux.

Using the corrected photometry, we construct a likelihood function of the proposal vector $\theta = \{T_\text{eff}, R, d, A_v\}$ for each white dwarf in the sample:
\begin{align}
    \ln &\mathcal{L}(T_\text{eff}, \log g, d, A_v) =  \\
    &-\frac{1}{2}\sum_{i=1}^n\left[\frac{(f_{i,\text{obs}} - f_{i}(\theta))^2}{\sigma_{f_i}^2} + \ln 2\pi\sigma_{f_i}^2\right] \nonumber \\ 
    &- \frac{1}{2}\left[ \frac{(d - 1/\varpi_\text{obs})^2}{\sigma_{\varpi_\text{obs}}^2} + \ln 2\pi\sigma_{\varpi_\text{obs}}^2\right] + 2\ln d  \nonumber \\
    &-\frac{1}{2}\left[\frac{(A_V - A_{V\text{,obs}})^2}{(0.1A_{V\text{,obs}})^2} + \ln 2\pi (0.1A_{V\text{,obs}})^2\right] \nonumber
\end{align}
where $\varpi_\text{obs}$ is the parallax of the main sequence companion as measured by Gaia for the wide binaries and the parallax of the white dwarf for the single stars. We apply zero point corrections to the parallax as reported by \cite{2021MNRAS.508.3877G} and we explore the parameter space using \texttt{emcee} \citep{2013PASP..125..306F} to construct posterior distributions for all parameters in the proposal vector. For each star, we calculate mean vectors and covariance matrices from the $T_\text{eff}$ and radius posterior samples.

We validate the measured radii of the isolated white dwarf sample against the photometric radius measurements of \cite{2021MNRAS.508.3877G} and the spectroscopic radius measurements of \cite{2009A&A...505..441K}. We find a bias in our results relative to \cite{2021MNRAS.508.3877G} of $1.2\%$, and $5.2\%$ relative to \cite{2009A&A...505..441K} (that is our measured radii are systematically smaller than the referenced datasets by $1.2\%$ and $5.2\%$). This is consistent with the known offsets between photometric and spectroscopic measurements of  \cite{2019ApJ...871..169G}, who found that effective temperatures from spectroscopy and photometry could diverge by as much as $10\%$. Indeed we find that the photometric benchmark dataset shows a bias of $5.1\%$ relative to the spectroscopic benchmark in radius. Figure \ref{fig:benchmark} demonstrates our agreement with external datasets. Our measurement of effective temperature has a $1.7\%$ bias relative to \cite{2021MNRAS.508.3877G} and $2.4\%$ relative to \cite{2009A&A...505..441K}.

\subsection{Gravitational Redshift}

For the white dwarfs with main sequence wide companions, we adopt the average gravitational redshift measurements from the H$\alpha$ line and the H$\alpha$ and H$\beta$ lines reported by \cite{2025A&A...695A.131R}, weighted by the inverse square of their uncertainties. They measure gravitational redshift from white dwarf radial velocity by subtracting out the Gaia-reported radial velocity of its main-sequence companion and adding their uncertainties in quadrature, with corrections for the small gravitational redshift of the main sequence companion, which is typically of order $0.1$~km\,s$^{-1}$.

For white dwarfs from the isolated sample, gravitational redshifts cannot be directly measured because the star's true space motion cannot be removed from the measurement. To infer the gravitational redshift-radius relation we therefore must use a statistical method to marginalize over the distribution of true space motions, which we describe in the subsequent section. We adopt the radial velocities measured from the non-local thermodynamic equilibrium core of the H$\alpha$ and H$\beta$ lines by \cite{2025moi} and corrected to the reference frame of the local standard of rest, which they show to be in good agreement with previous analyses.

\begin{figure}
    \centering
    \includegraphics[width=\linewidth]{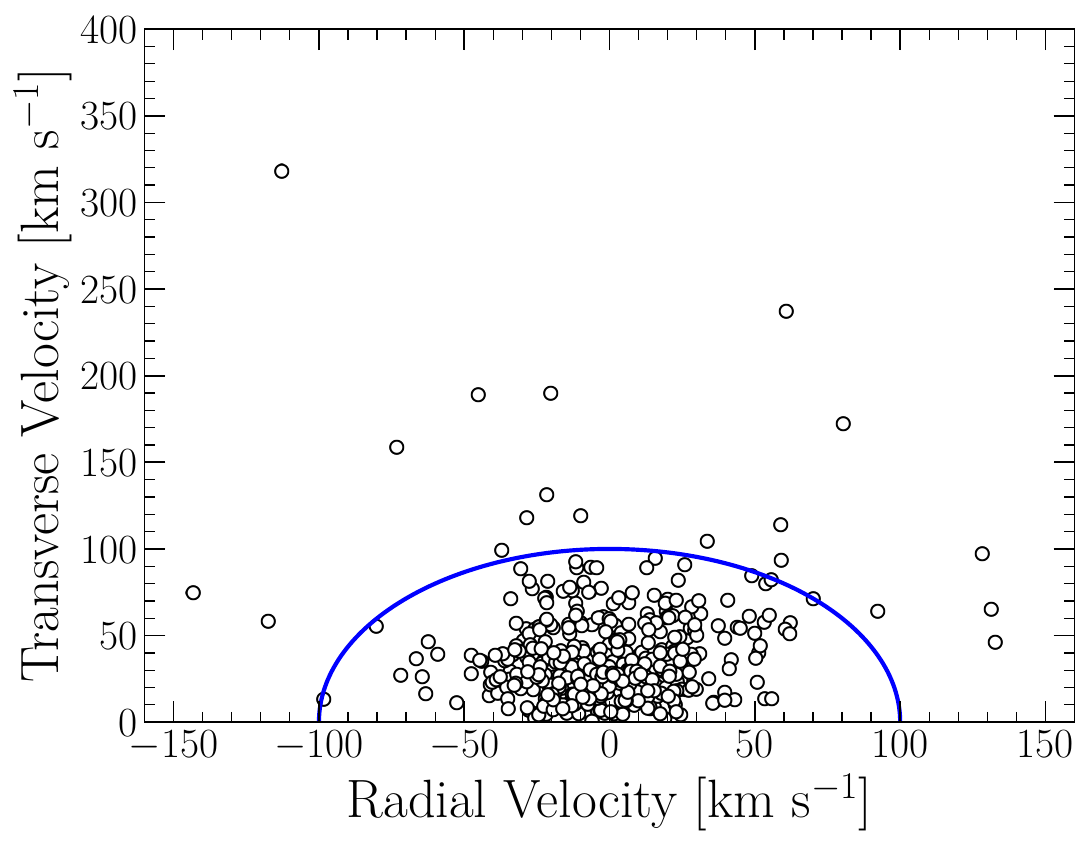}
    \caption{Kinematics of the isolated white dwarf sample. Stars with total three dimensional velocities greater than $100$~km\,s$^{-1}$ (indicated by the blue circle) are assumed to have kinematics inconsistent with the thin disk, and are therefore removed from the sample. This removes 13 objects from the analysis.}
    \label{fig:toomre}
\end{figure}

Our analysis procedure requires that the radial velocities of the isolated stars in our sample be drawn from the same distribution. This means that we must remove objects with non-thin disk kinematics from our sample. First, we perform a $3\sigma$ clip on the radial velocity distribution, which removes 11 stars from the sample. Following \cite{2025moi}, we construct a Toomre diagram, plotting transverse velocities from Gaia against the radial space motion of the star. The latter quantity is estimated as the measured radial velocity subtracted by the theoretical gravitational redshift estimated from the star's estimated temperature and radius. We remove 13 stars from the sample with total three-dimensional velocities greater than $100$~km\,s$^{-1}$. Figure \ref{fig:toomre} presents the velocity distribution of the sample. Our choice of $100$~km\,s$^{-1}$ is somewhat arbitrary, although the sample is insensitive to the exact value chosen: selecting objects with velocities less than $80$~km\,s$^{-1}$ removes only 21 additional objects and does not substantially effect the results of our analysis.

\subsection{Properties of the Sample}


\begin{figure*}
    \centering
    \includegraphics[width=\linewidth]{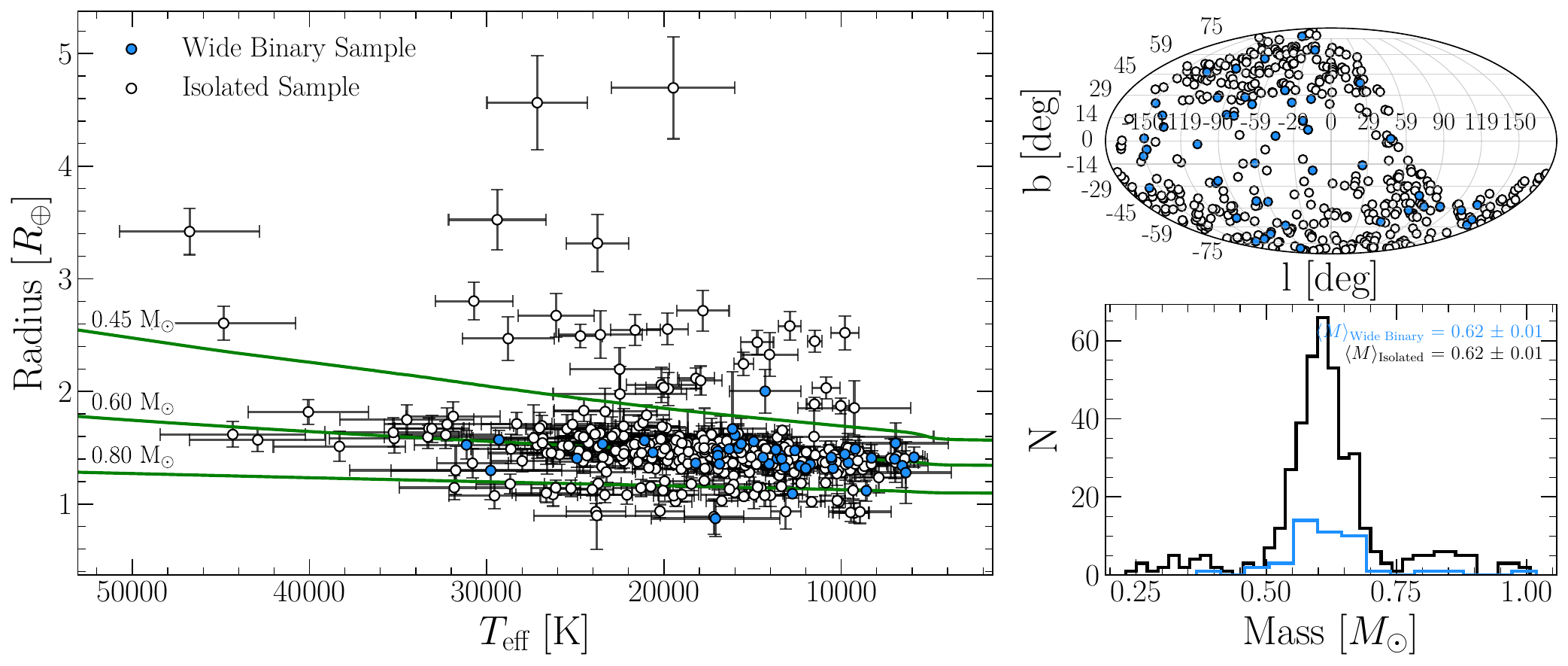}
    \caption{Measured parameters for the wide binary and isolated samples, as well as on-sky distributions and ages inferred from MIST. We mark the theoretical temperature-radius curves for a cooling white dwarf of masses $0.4~M_\odot$, $0.6~M_\odot$, and $0.8~M_\odot$ using the mass-radius relation of \cite{2020ApJ...901...93B}, and the mass distributions are determined using the same relations. Those models assume a core composition of carbon and oxygen in equal ratios, meaning that the distribution masses of objects below $\approx 0.45 M_\odot$, which may have helium cores, is not necessarily accurate. Our targets are biased to the southern sky due to the location of the VLT, making our corrections to the local standard of rest important for accurate inference.}
    \label{fig:sample}
\end{figure*}

We append the measured gravitational redshift or radial velocity of each star to the previously constructed vectors of mean effective temperature and radius, and we add the variance of the gravitational redshift or radial velocity to the covariance matrix. We assume that gravitational redshifts and radial velocities are independent from radius and effective temperature. 
Our final sample size consists of $468$ white dwarfs. The measured properties of our samples are presented in Figure \ref{fig:sample}, as well as their on-sky distribution. In principle, there also exists a sinusoidal variation in radial velocity as a function of galactic longitude in our isolated white dwarf sample due to the rotation of spiral arms of the galaxy. Our sampling of galactic longitudes is sufficient that this is not a significant source of scatter in our analysis, as the variations will average to zero. 

\section{Hydrogen Envelope Thickness} \label{sec:deconv}

The task of inferring a representative hydrogen layer from our mean value vectors covariance matrices is fundamentally the problem of inferring the underlying distribution from which a noisy set of data is drawn. Given that we expect the noise distribution from our data to be heteroskedastic (e.g. higher mass white dwarfs are likely to have higher uncertainties on account of their intrinsic faintness), this is a nontrivial problem. 

We infer the underlying distribution via extreme deconvolution, following the methodology established by \cite{2011AnApS...5.1657B}. We model the probability of the true radius, effective temperature, and gravitational redshift values $\text{\bf{v}}$ using a Gaussian mixture model:
\begin{equation}
    \text{p}(\text{{\bf v}}) = \sum_{i=1}^k\alpha_i\mathcal{N}(\text{{\bf v}} | \bm{\mu}_i, \bm{V}_i)
\end{equation}
with amplitudes $\alpha_i$ that sum to unity, and where the components of the mean vector are $\bm{\mu}_i = (R, T_\text{eff}, v_\text{g})^\top$.

For the white dwarfs with wide binary companions for which gravitational redshifts can be directly measured, we directly infer parameters $\theta = (\alpha_i, \bm{\mu}_i, \bm{V}_i)$ from the observed measurements. For white dwarfs without wide binary companions, gravitational redshift cannot be directly measured. Assuming that the noise present in our measurements is unbiased Gaussian noise, the measured radial velocity consists of
\begin{equation}
    v_\text{obs} = v_\text{g} + \epsilon + \nu
\end{equation}
where $\epsilon \sim \mathcal{N}(0,\sigma^2_\text{err})$ represents the measurement noise term and $\nu \sim \mathcal{N}(0, \sigma^2_s)$ represents the true space motion of the star. This expression can be rewritten in terms of a new parameter $\delta = \epsilon + \nu \sim \mathcal{N}(0, \sigma^2_\text{err} + \sigma^2_\text{s})$ as
\begin{equation}
    v_\text{obs} = v_\text{g} + \delta
\end{equation}
and so we account for the true space motion of the star by adding the variance of the true space velocity distribution to the diagonal element of the covariance matrix representing the variance of the radial velocity measurement. We estimate the variance due to the real space motion of the star by subtracting in quadrature the standard deviation of the wide binary gravitational redshifts from the standard deviation of the isolated white dwarf radial velocities. Our adopted value is $\sigma = 24.4$~km\,s$^{-1}$, which is consistent with the velocity dispersions of \cite{2022A&A...658A..22R} for a relatively young population (cf. the age distribution of the SPY sample, our isolated white dwarf sample, measured in that analysis).

The parameters of interest in our system span several orders of magnitude (radii are on the order of $0.01$~$R_\odot$ whereas $T_\text{eff}$ are on the order of $10^3$~K). Therefore, prior to inference we regularize our observations $\bm{x}_j$ of mean radius, temperature, and gravitational redshift as well as the associated covariance matrices $\Sigma_{X,j}$ via the linear transformation
\begin{align} \label{eq:transform}
    \bm{x}_j &\to \text{A}(\bm{x}_j - \overline{\bm{x}}) \nonumber \\
    \Sigma_{X,j} &\to \text{A}\Sigma_{X,j} \text{A}^\top 
\end{align}
Where $\text{A}$ is the diagonal matrix whose elements are the inverse of the sample variance of $\bm{x}_j$, and $\overline{\bm{x}}$ is the sample mean vector. This transformation ensures that variations in observed quantities are at comparable scales while preserving their covariances.

We then infer the parameters of the Gaussian mixtures for both the regularized wide binary and single white dwarf observations using the expectation maximization algorithm \citep{2011AnApS...5.1657B}. Finally, we transform the parameters $\bm{\mu}_i$ and $\bm{V}_i$ of the inferred regularized Gaussians back into radius, velocity, and effective temperature via the inverse of the transformation specified in Equation \ref{eq:transform}.

\begin{figure}
    \centering
    \includegraphics[width=\linewidth]{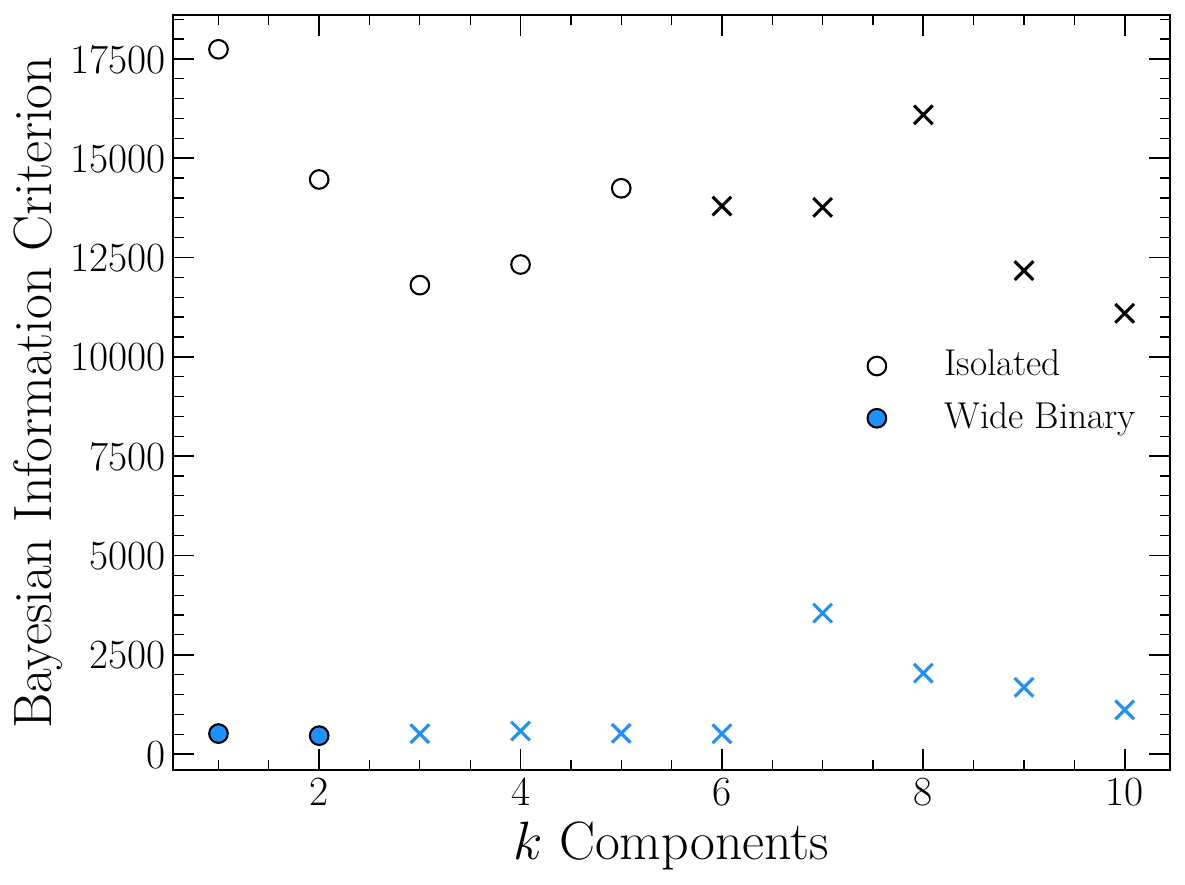}
    \caption{Logarithm of Bayesian information criterion as a function of number of Gaussian components for deconvolution of the observed mass-radius relation (e.g. \citealt{2011AnApS...5.1657B}). Component numbers which result in singular matrices are plotted marked as ``x''. This is a particular issue for the wide binary sample, due to its much smaller sample size. The score of the non-singular fits are similar to those of the singular fits though, indicating that the information is still captured. The isolated sample is best represented by three components, and the wide binary sample by two.}
    \label{fig:bic-scatter}
\end{figure}

\begin{figure*}
\centering
    \includegraphics[width=\linewidth]{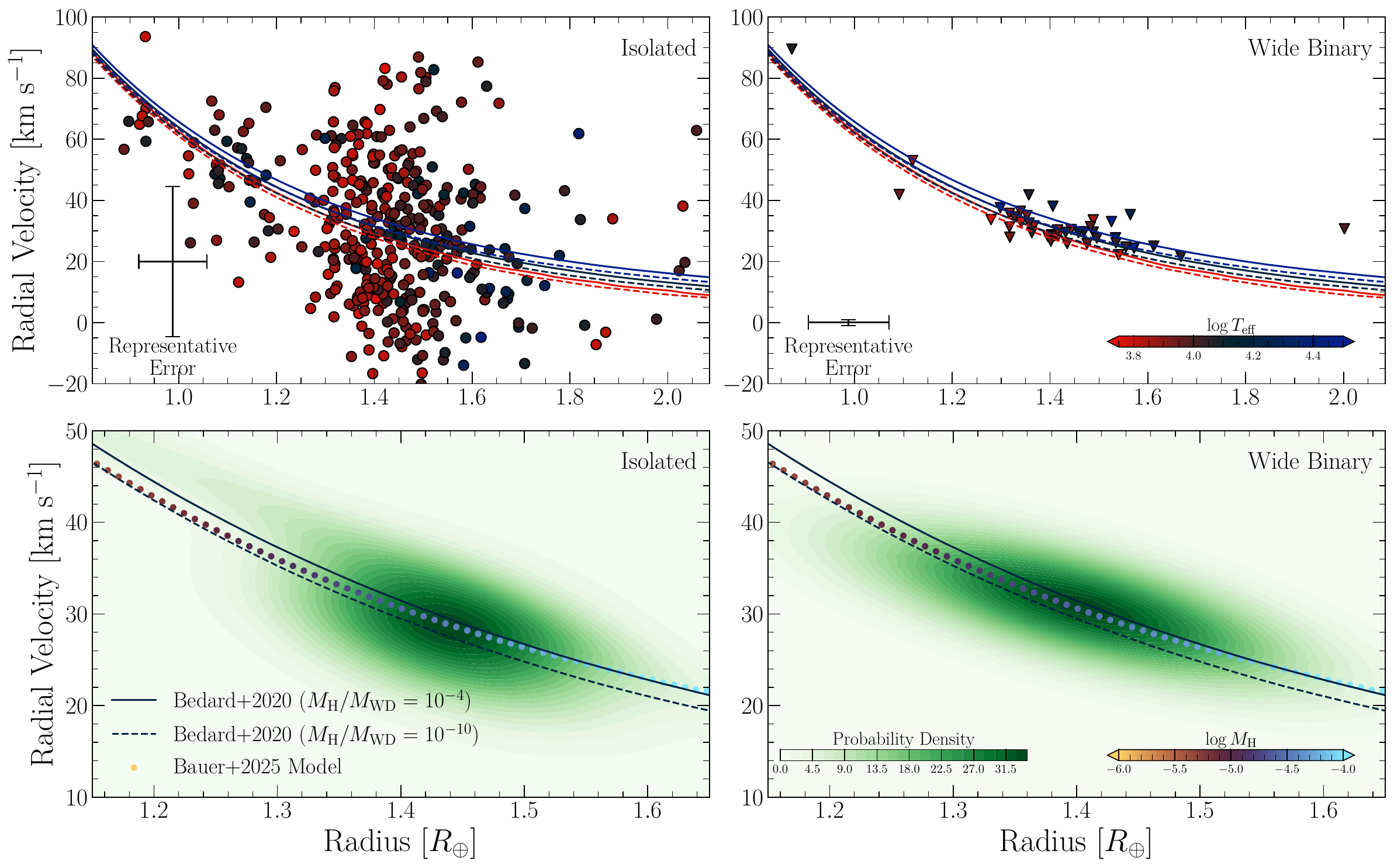}
    \caption{(\textit{Top}) The gravitational redshift-radius relation colored by effective temperature of the isolated white dwarf sample (\textit{left}) and the wide binary sample of \cite{2025A&A...695A.131R} (\textit{right}). We plot theoretical relations for thick hydrogen envelopes of $M_\text{H}/M_\text{WD} = 10^{-4}$ in solid lines and thin hydrogen envelopes of  $M_\text{H}/M_\text{WD} = 10^{-10}$ in dashed lines  \citep{2020ApJ...901...93B}. (\textit{Bottom}) Optimal probability density function, deconvolved from uncertainties and marginalized over effective temperature, for isolated white dwarfs (\textit{left}) and the wide binary sample (\textit{right}). We plot theoretical relations for the expected effective temperature of the given probability density function. Additionally, the mass-radius relation inferred from MIST isochrones \citep{2025arXiv250921717B} is shown, colored by the hydrogen envelope mass assumed at each point.}
    \label{fig:relation}
\end{figure*}

We select $k$, the number of Gaussian components to use in the deconvolution, by testing a range of values and selecting the one which minimizes the Bayesian information criterion relative to the data. In particular, we test using $k$ values of one through nine. We choose to use Bayesian information criterion as our goodness-of-fit metric because it penalizes both underfitting and overfitting (the latter being of greatest concern for this methodology). 

Fitting too many Gaussian components can lead the expectation maximization algorithm to attempt to model individual datapoints using their own unique mixture element. This can result in an ill-defined distribution, generating a mixture component with a singular covariance matrix. This is a problem for the wide binary dataset in particular, given the small sample size. We take two steps to ensure that the inferred distribution is well-behaved: first, we apply a $3\sigma$ cut on the wide binary gravitational redshift and radius distributions to filter outliers; this removes three datapoints. Second, we calculate the determinant of all mixture components' covariance matrices at each step $k$ for both datasets. If any covariance matrix is found to be singular, we reject that value of $k$. The results of our search are presented in Figure \ref{fig:bic-scatter}. We find that the wide binary dataset is best represented by two Gaussian components, and the isolated sample by three. The wide binary sample is particularly affected by singular matrices, which is consistent with its smaller sample size. This technique leaves us with probability density functions of radius, effective temperature, and gravitational redshift for the isolated white dwarf data and the wide binary data.

\section{Results} \label{sec:results}

\subsection{Hydrogen Envelope Thickness}

The observed gravitational redshift-radius relations and deconvolved probability density functions marginalized over effective temperature are displayed in Figure \ref{fig:relation}. We find that both the isolated white dwarf and wide binary samples yield similar probability density functions. For the isolated white dwarf sample, the expectation maximization algorithm was found to be insensitive to data cleaning procedures such as $3\sigma$ clipping the radial velocity distribution and the choice of $100$~km\,s$^{-1}$ as the total three dimensional velocity cut, indicating that the parameters inferred are robust to outliers.  

We plot theoretical mass-radius relations assuming a constant thick hydrogen envelope of $M_\text{H}/M_\text{WD}=10^{-4}$ \citep{2020ApJ...901...93B}, as well as the mass-radius relation predicted by the same code with a thin hydrogen envelope fixed at $M_\text{H}/M_\text{WD}=10^{-10}$. Additionally, we plot the mass-radius relation derived from the extension of the MESA Isochrones and Stellar Tracks\footnote{\url{https://mist.science/}} (MIST; \citealt{2016ApJS..222....8D, 2016ApJ...823..102C, 2025arXiv250921717B, mist_dotter}) into the white dwarf cooling sequence. For these models, the mass of the white dwarf hydrogen envelope is determined by the evolutionary history of the star. This yields a decreasing hydrogen envelope mass with increasing stellar mass due to increased residual thermonuclear burning \citep{2019MNRAS.484.2711R}. We adopt a solar metallicity white dwarf progenitor ($[\text{Fe}/\text{H}]=[\alpha/\text{Fe}]=0$) for the MIST models in order to most accurately compare with other models.

\begin{figure}
    \centering
    \includegraphics[width=\linewidth]{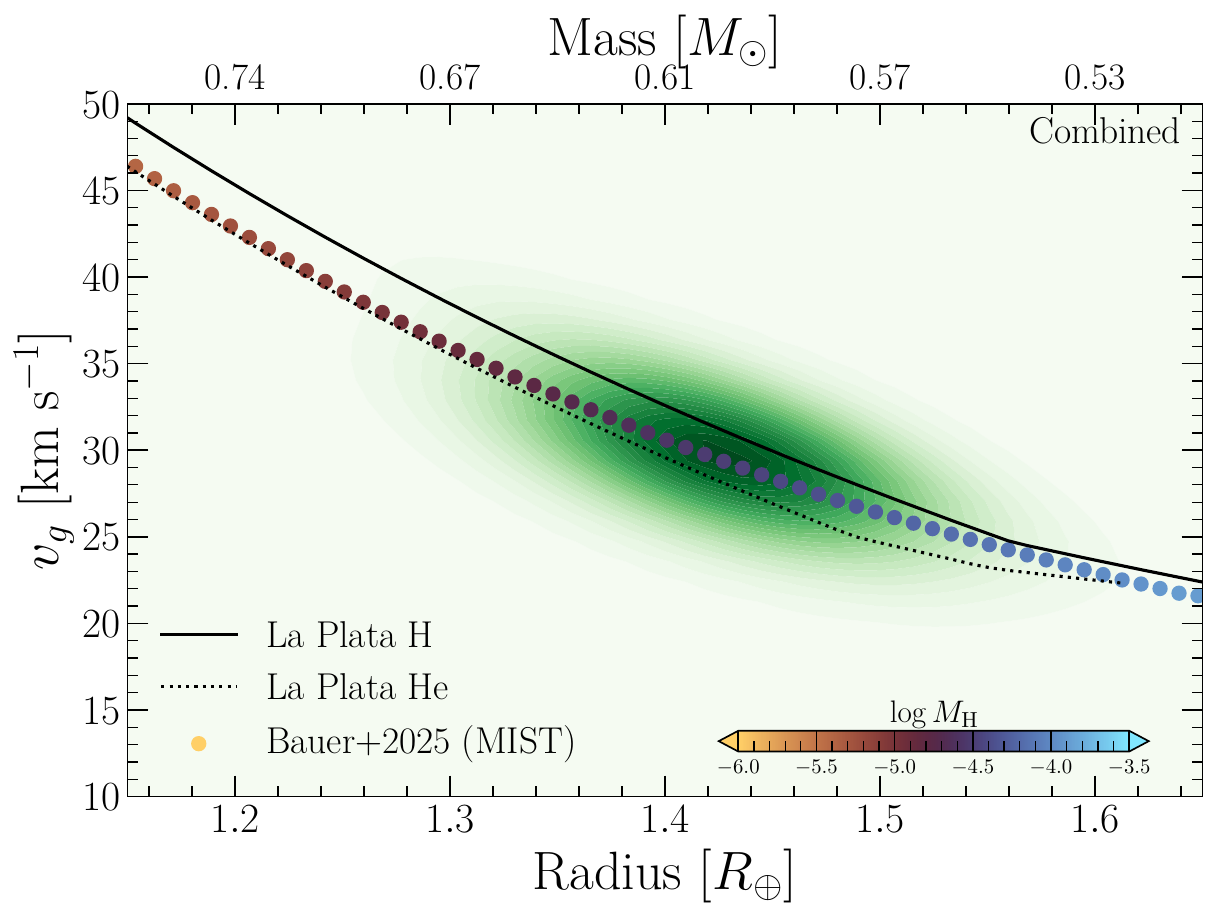}
    \caption{The combined probability density function for all white dwarfs in the sample plotted against La Plata theoretical models for a thick hydrogen envelope (solid) and a helium envelope (dashed; \citealt{2016ApJ...823..158C, 2017ApJ...839...11C}), and for the mass-radius relation inferred from MIST isochrones \citep{2025arXiv250921717B}. Like the MIST models, the thick La Plata model finds a hydrogen layer mass which decreases as a function of stellar mass, however they find systematically more massive hydrogen layers due to different treatment of microphysics. Masses are interpolated from the MIST relation assuming a temperature of $16,100$~K, the mean mass of the sample.}
    \label{fig:combined}
\end{figure}

\begin{deluxetable}{ccccc}
\label{tab:params}
\tablecaption{Bayes factors between the fixed hydrogen layer models of \cite{2020ApJ...901...93B} and MIST, as well as the La Plata models and MIST, for the wide binary sample, the isolated white dwarf sample, and the combined sample. Our results indicate preference for the MIST and La Plata thick models, although the statistical power is not yet great enough to conclusively discriminate between models.}
\tablewidth{1300pt}
\tabletypesize{\scriptsize}
\tablehead{Sample & \colhead{$M_\text{H}/M_\text{WD} = 10^{-4}$} & \colhead{$M_\text{H}/M_\text{WD} = 10^{-10}$} & \colhead{La Plata H} & \colhead{La Plata He}}
\startdata
Wide Binary & 1.22 & 1.30 & 1.14 & 1.38 \\ 
Isolated & 1.29 & 1.38 & 1.18 & 1.59 \\ 
Combined & 1.16 & 1.24 & 1.12 & 1.25
\enddata
\end{deluxetable}

Because our data are independent and (in principle) independently distributed, we can combine the inferred distribution. Figure \ref{fig:combined} presents the combined probability distribution. Here, we plot the MIST models as well as the La Plata mass-radius relations assuming a thick hydrogen layer and a helium-only layer \citep{2016ApJ...823..158C, 2017ApJ...839...11C}. For these models, thick hydrogen layers are produced using full stellar evolution sequences; the hydrogen layers predicted by these models are greater than those of the MIST models, starting at $M_\text{H}/M_\text{WD}=10^{-3.5}$ for low-mass objects ($M_\text{WD} = 0.528$~$M_\odot$) and decreasing to $M_\text{H}/M_\text{WD}=10^{-5.2}$ for higher masses ($M_\text{WD} = 0.833$~$M_\odot$). The helium-only layers are then created by artificially stripping the hydrogen layer until the outer layer of the white dwarf is nearly pure helium. The masses of the La Plata evolutionary model calculations in this region are widely spaced, and so our models contain non-negligible linear interpolation error.

We find excellent agreement between our data and theoretical mass-radius relations. We quantify the model preference our data with Bayes factors relative to the MIST mass-radius model, presented in Table \ref{tab:params}. This is the ratio of the probability of the MIST model to the probability of other models given the data. We find that in the MIST models are best performing in all three of the wide binary, isolated white dwarf, and combined samples, followed by the La Plata thick models. The helium-only La Plata models are always least preferred models. This is consistent with expectations, since the helium-only La Plata models are produced by stripping all hydrogen from the evolved star and therefore should have the smallest radius. In general we find a preference for models with thick hydrogen layers which vary as a function of mass over those with fixed hydrogen layers, in keeping with the expected residual burning trends. The preference is slight and we do not have sufficient statistical power to conclusively rule out any mass-radius relation, however it is consistent across all samples.

\begin{figure*}
    \centering
    \includegraphics[width=\linewidth]{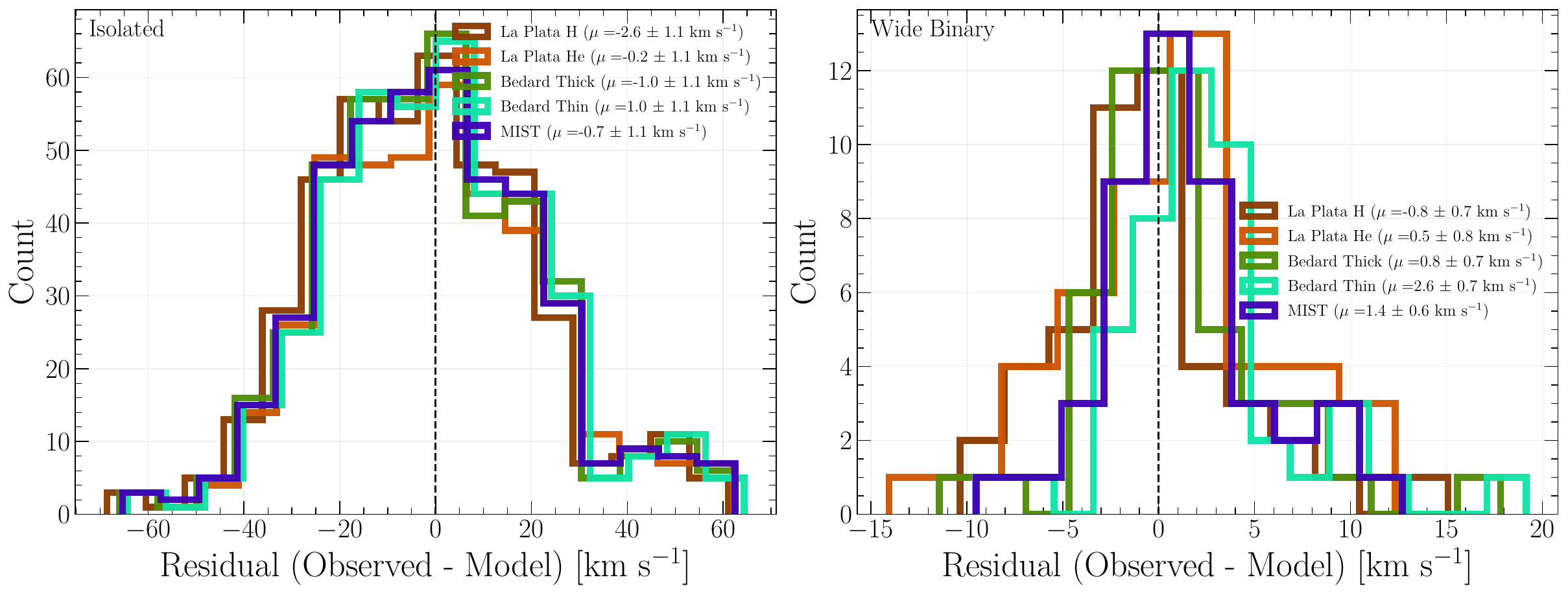}
    \caption{Residuals (observed minus expected) of gravitational redshifts inferred from models for the hydrogen and helium mass-radius models of \cite{2016ApJ...823..158C, 2017ApJ...839...11C}, the models of \cite{2020ApJ...901...93B} which assume constant hydrogen envelope masses of $M_\text{H}/M_\text{WD} = 10^{-4}$ and $M_\text{H}/M_\text{WD} = 10^{-10}$, and the MIST mass-radius relation \citep{2025arXiv250921717B}. On the left is the isolated white dwarf dataset, and the right is the wide binary dataset. The intrinsic scatter of the isolated white dwarf dataset is much greater than that of the wide binary dataset. The zero residual lines for both models are marked as dashed lines. We quantify the agreement of our inferred probability distributions with models in Table \ref{tab:params}.}
    \label{fig:residuals}
\end{figure*}

Figure \ref{fig:residuals} presents the residuals of each data point in our sample with respect to all five theoretical mass-radius models. We estimate the residuals by Monte Carlo sampling our inferred temperature and radius measurements to construct theoretical gravitational redshifts with realistic errors, which we add in quadrature with our observed radial velocity measurements. There, we find that the strongest constraints come from the wide binary dataset, the fixed thick model of \cite{2020ApJ...901...93B} and the La Plata models perform similarly. In the isolated white dwarf sample almost all observed values are smaller than the model predictions. This is likely driven in part by Eddington bias: large uncertainties in radius can flatten the mass-radius relation inferred from data, leading to smaller than anticipated residuals. The relatively sharp distribution of masses in the wide binary sample due to the smaller sample size mitigates this bias. 

\subsection{$^{12}$C($\alpha$, $\gamma$)$^{16}$O Reaction Rate}

Most of the white dwarfs in our sample have masses such that their cores should be composed principally of $^{12}$C and $^{16}O$ (though objects with mass $<0.45~M_\odot$ will likely have He cores). Uncertainties in the $^{12}$C($\alpha$, $\gamma$)$^{16}$O reaction rate will alter their core composition to a degree which may be significant for this analysis: a higher reaction rate creates a core which is richer in $^{16}$O. The resulting ion-ion interaction terms cause a second-order perturbation to the equation of state which causes the star to have a more compact structure, increasing its gravitational redshift \citep{2019MNRAS.490.5839B}.

We test this by calculating theoretical gravitational redshifts from the cooling sequence of a $0.565$~$M_\odot$ white dwarf as calculated by \cite{2023ApJ...954...51C}. These cooling sequences are calculated with $^{12}$C($\alpha$, $\gamma$)$^{16}$O reaction rates drawn from from $-3\sigma$ to $+3\sigma$ from the experimentally inferred probability distribution function in steps of $0.5\sigma$ \citep{2017RvMP...89c5007D, 2020PhRvL.125r2701K, 2022JPhG...49k0502S}. We find that variations in the $^{12}$C($\alpha$, $\gamma$)$^{16}$O reaction rate result in changes to a $0.565~M_\odot$ model white dwarf's gravitational redshift on the order of $0.1$~km s$^{-1}$. The signal is not likely detectable in our sample.

\section{Conclusions} \label{sec:conclusions}

We have presented a highly accurate measurement of the white dwarf mass radius relation using gravitational redshifts. Using a Gaussian mixture model allows us to infer a continuous probability distribution which can be used for robust hypothesis testing. We find that the MIST mass-radius relation calculated by \cite{2025arXiv250921717B} is the most consistent with the observed probability distribution, followed by the La Plata mass-radius relation assuming a thick hydrogen layer of \cite{2016ApJ...823..158C}. Generally we find that models which account for the evolution of the progenitor star, implying a hydrogen layer mass which decreases with increasing stellar mass, better reproduce the observed distribution than those assuming a fixed hydrogen layer mass. Our results are consistent across independent samples, but we do not have sufficient statistical power to conclusively reject any model.

The point of greatest leverage for improvement of this analysis is the wide binary sample size. Our wide binary dataset outperforms the isolated white dwarf dataset, despite having only one tenth the sample size. Compared to the potential sample size available from low-resolution surveys such as the Sloan Digital Sky Survey or the Dark Energy Survey, our analysis is extremely small. If the systematic effects present in the low-resolution data can be accurately characterized, or if more high-resolution spectroscopy becomes available for white dwarfs in wide binaries, it will likely be possible to exploit the disproportionately large contribution of the hydrogen envelope to the radius of white dwarfs to make precise measurements of hydrogen envelope mass using gravitational redshift measurements.

\section*{Data Availability}

Our data, as well as the code needed to reproduce our results, are available at \url{https://github.com/stefanarseneau/hydrogen-layer} and are preserved on Zenodo at \url{https://doi.org/10.5281/zenodo.18652205}.


\begin{acknowledgments}

We thank the anonymous referee for the constructive feedback, which improved the quality of this manuscript. MEC acknowledges support from grant RYC2021-032721-I and RR acknowledges grant RYC2021-030837-I, both funded by MCIN/AEI/ 10.13039/501100011033 and by “European Union NextGeneration EU/PRTR”. This research was partially supported by the AGAUR/Generalitat de Catalunya grant SGR-386/2021 and the Spanish MINECO grant, PID2023-148661NB-I00. SMA and JJH acknowledge funding in-part provided by HST Program No. 16719. Work by EB was performed under the auspices of the U.S. Department of Energy by Lawrence Livermore National Laboratory under Contract DE-AC52-07NA27344.

This work has made use of data from the European Space Agency (ESA) mission Gaia (\url{https://www.cosmos.esa.int/gaia}), processed by the Gaia Data Processing and Analysis Consortium (DPAC, \url{https://www.cosmos.esa.int/web/gaia/dpac/consortium}). Funding for the DPAC has been provided by national institutions, in particular the institutions participating in the Gaia Multilateral Agreement.

This research has made use of the VizieR catalogue access tool, CDS, Strasbourg, France \citep{10.26093/cds/vizier}. The original description of the VizieR service was published in \citet{vizier2000}. Based on data obtained from the ESO Science Archive Facility with DOI(s): \url{https://doi.eso.org/10.18727/archive/50}. This research has made use of NASA Astrophysics Data System.

\end{acknowledgments}

\bibliography{citations}{}
\bibliographystyle{aasjournalv7}



\end{document}